\documentclass[12pt]{article}
\pdfoutput=1
\usepackage{graphicx}
\usepackage{amsfonts}
\usepackage{amsmath}
\usepackage[mathscr]{eucal}
\usepackage{booktabs}

\def\href#1#2{#2}   
\newif\ifdraft
\draftfalse	  
\reversemarginpar   
\makeatletter
\let\mlabel=\label
\let\adkendequation=\endequation%
\def\endequation{\adkendequation\adklabel\global\@ignoretrue}
\let\adkendeqnarray=\endeqnarray%
\def\endeqnarray{\adkendeqnarray\adklabel\global\@ignoretrue}
\newbox\marglabbox
\def\adklabel{\ifvoid\marglabbox\else\marginpar{\unhbox\marglabbox}\fi}
\def\label#1{\ifdraft\ifmmode%
  \global\setbox\marglabbox=\hbox{\hfill\fbox{\tiny\verb*~#1~}}%
  \else\ifinner\else\marginpar{\hfill\fbox{\tiny\verb*~#1~}}%
  \fi\fi\fi \mlabel{#1}}
\makeatother
\ifdraft%
\fi
\def\eusm{\mathscr}
\font\twelvefrak=eufm10 scaled 1200
\font\tenfrak=eufm10
  \newfam\frakfam
  
  \textfont\frakfam=\twelvefrak
  \scriptfont\frakfam=\tenfrak
  \scriptscriptfont\frakfam=\scriptfont\frakfam
\def\sqr#1#2{{\vcenter{\hrule height.#2pt
   \hbox{\vrule width.#2pt height#1pt \kern#1pt
      \vrule width.#2pt}
   \hrule height.#2pt}}}

\def\bsqr#1#2{{\vrule width #1pt height#2pt}}
\def\bsquare{{\mathchoice\bsqr66\bsqr66\bsqr33\bsqr33}}
\def\badbreak{\penalty1000}

\begin{document}

\begin{center}
{\Large{\bf Phases of SU(3) Gauge Theories with Fundamental}} \\
\vspace*{.14in}
{\Large{\bf Quarks via Dirac Spectral Density}} \\
\vspace*{.24in}
{\large{Andrei Alexandru$^1$ and Ivan Horv\'ath$^2$}}\\
\vspace*{.24in}
$^1$The George Washington University, Washington, DC, USA\\
$^2$University of Kentucky, Lexington, KY, USA


\end{center}

\vspace*{0.05in}

\begin{abstract}

\noindent
We propose that, in SU(3) gauge theories with fundamental quarks, confinement 
can be inferred from spectral density of the Dirac operator. This stems from 
the proposition that its possible behaviors are exhausted by three distinct 
types (Fig.1). The monotonic cases are standard and entail confinement with 
valence chiral symmetry breaking (A) or the lack of both (C,C'). The bimodal 
({\em anomalous}) option (B) was frequently regarded as an artifact (lattice or 
other) in previous studies, but we show for the first time that it persists in 
the continuum limit, and conclude that it informs of a non-confining phase 
with broken valence chiral symmetry. This generalization rests on the following. 
$(\alpha)$~We show that bimodality in N$_f$=0 theory past deconfinement 
temperature $T_c$ is stable with respect to removal of both infrared and 
ultraviolet cutoffs, indicating that anomalous phase is not an artifact.
$(\beta)$ We demonstrate that transition to bimodality in N$_f$=0 is simultaneous
with the loss of confinement: anomalous phase occurs for $T_c \!<\! T \!<\! T_{ch}$, 
where $T_{ch}$ is the valence chiral restoration temperature.
$(\gamma)$ Evidence is presented for thermal anomalous phase in N$_f$=2+1 QCD at 
{\em physical} quark masses, whose onset too coincides with the conventional 
``crossover $T_c$''. We conclude that the anomalous regime 
$T_c \!<\! T \!<\! T_{ch}$ is very likely a feature of nature's strong 
interactions. 
$(\delta)$~Our past studies of zero-temperature N$_f$=12 theories revealed that 
bimodality also arises via purely light-quark effects. As a result, we 
expect to encounter anomalous phase on generic paths to valence chiral 
restoration. We predict its existence also for $N_f$ massless flavors ($T\!=\!0$) 
in the range $\mbox{\rm N}_f^c \!<\! \mbox{\rm N}_f \!<\! \mbox{\rm N}_f^{ch}$, 
where N$_f^c$ could be quite low. Conventional arguments would associate 
$\mbox{\rm N}_f^{ch}$ with the onset of conformal window.

\end{abstract}

\vspace*{0.10in}

\noindent{\bf 1. Introduction.}
SU(3) gauge theories with quarks in fundamental representation are important 
in elementary particle physics. Indeed, the strong dynamics of ``real-world'' 
quarks and gluons is believed to be described within this context, and 
the near-conformal infrared behavior in theories with many massless flavors 
(just below conformal window) may be relevant for theories of technicolor type. 
In addition, the systems within conformal window are attractive test beds of 
conformal dynamics in four dimensional quantum field theory. 

Since the quarks of nature are not massless, it is desirable to
consider all theories of the above type, i.e. including any number 
of arbitrarily massive quarks. Denoting such theory space at zero 
temperature as ${\eusm T}_0$ and at any temperature as 
${\eusm T} \supset {\eusm T}_0$, this wide landscape involves various
kinds of dynamics, expected to be well distinguished by corresponding vacuum 
properties.~\footnote{Note that, in what follows, we will not differentiate
between ``vacuum'' (zero--temperature concept) and 
``equilibrium state'' (finite--temperature concept). Path integral 
description of the state (via typical configurations) makes little 
difference between the two at the technical level.} In that regard, 
confinement and spontaneous chiral symmetry breaking (SChSB) dominate
the thinking about strongly interacting dynamics. Nevertheless, classifying 
elements of ${\eusm T}$ in these terms involves both conceptual 
and practical issues.

One problem is that SChSB is well-defined only when at least a pair of quarks 
is massless. However, the vacuum of any theory in ${\eusm T}$ can be probed 
for its ability to support long-range order via {\em valence} Goldstone pions 
(see e.g.~\cite{Ale12D, Ale14A}). Formally, a pair of valence quarks and 
a pair of action-compensating pseudofermions is added to 
the system~\cite{Mor87A}. In the massless valence limit, flavored chiral 
rotations of valence fields $\eta , \bar{\eta}$ are elevated to symmetries, 
which the vacuum either respects or not. The latter possibility entails valence 
spontaneous chiral symmetry breaking (vSChSB), indicated by non-zero value of 
$\langle \bar{\eta} \eta \rangle$. SChSB and vSChSB are identical notions 
whenever SChSB is meaningful, but vSChSB provides for a symmetry-based 
distinction among vacua over the whole set ${\eusm T}$.

The situation is different in case of confinement, whose meaning is quite 
intuitive throughout ${\eusm T}$, but a satisfactory consensus 
on its precise definition in such general context (even within 
${\eusm T}_0$) is lacking. Moreover, existing definitions are technically 
difficult to verify in a given theory. Convenient symmetry-based 
interpretation is only available when quarks do not affect vacuum 
correlations: in pure glue (N$_f$=0) theories at finite 
temperature~\cite{Pol78A,Sve82A}. 
In this restricted case, the breakdown of Z$_3$ symmetry, signaled by 
non--zero Polyakov line $\langle L  \rangle \neq 0$ or its magnitude 
$\langle |L|  \rangle > 0$, distinguishes confined and deconfined dynamics. 

\begin{figure}[t]
\begin{center}
    \centerline{
    \hskip 0.00in
    \includegraphics[width=4.4truecm,angle=0]{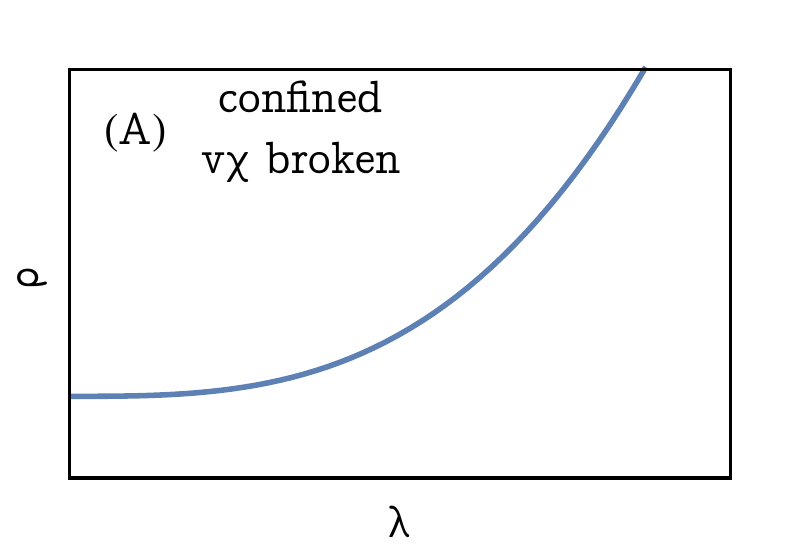}
    \hskip -0.15in
    \includegraphics[width=4.4truecm,angle=0]{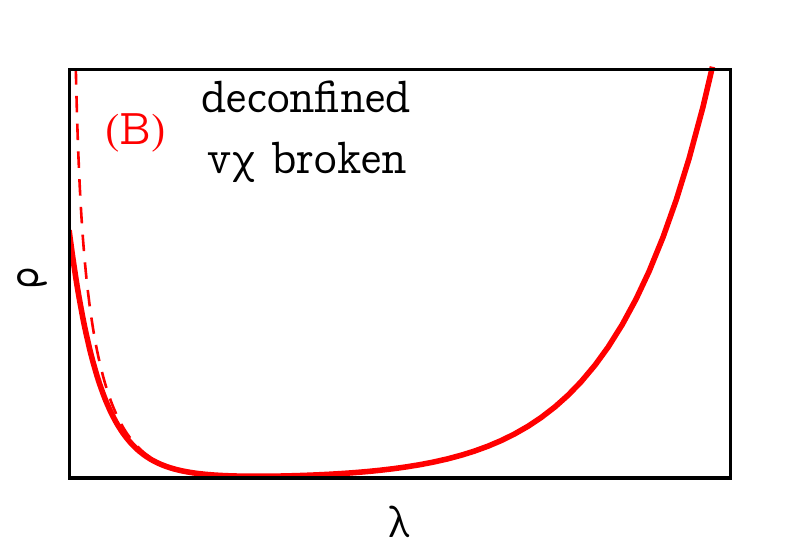}
    \hskip -0.15in
    \includegraphics[width=4.4truecm,angle=0]{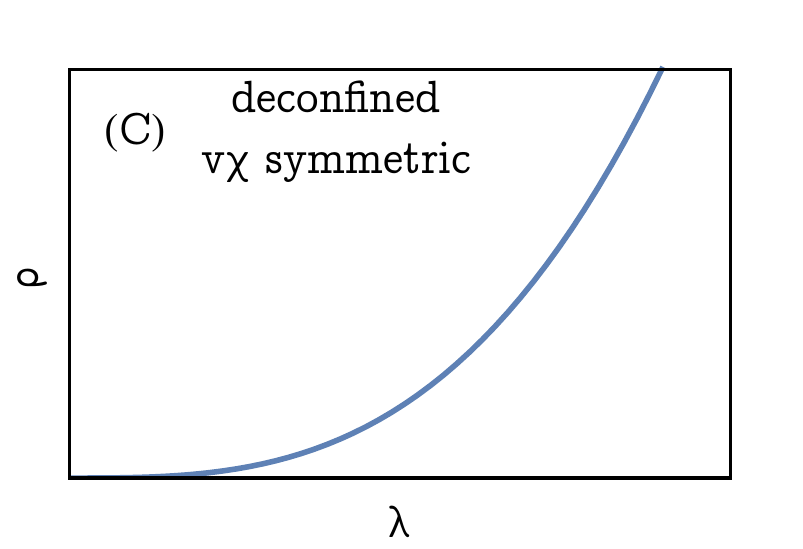}
    \hskip -0.15in
    \includegraphics[width=4.4truecm,angle=0]{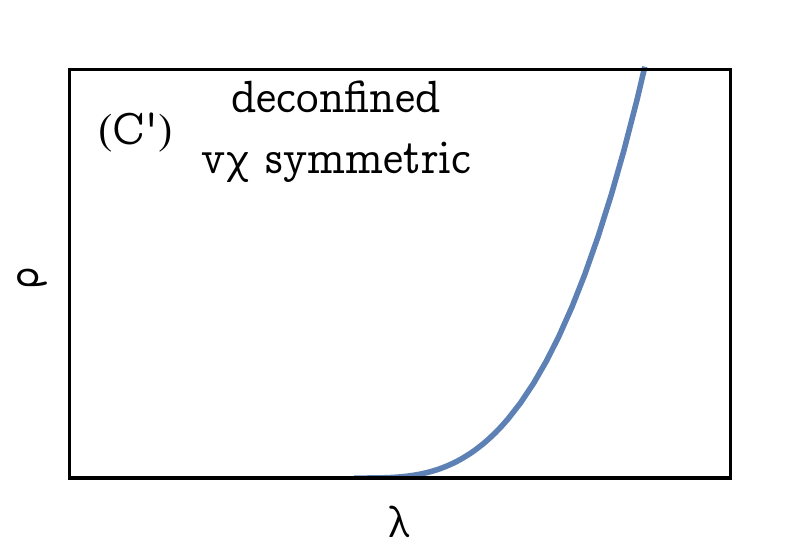}
     }
     \vskip 0.0in
     \caption{Confinement/vSChSB structure of ${\eusm T}$ via infrared
     behavior of Dirac spectral density.}
     \label{fig:illus1}
    \vskip -0.40in
\end{center}
\end{figure} 

Despite of issues with its formal definition over ${\eusm T}$, confinement 
can have simple and unambiguous dynamical signatures. Indeed, since 
the phenomenon shapes dynamics in a crucial manner, it is reasonable to expect 
that even simple vacuum observables are qualitatively affected. While 
details of this influence depend on the unknown specifics of confinement 
mechanism, the logic here is that of reverse engineering: 
if simple signatures can be empirically developed, consistently with existing 
results of lattice simulations and with theoretically clean N$_f$=0 case, 
then a valuable insight into the mechanism is obtained. In addition, practical 
benefits for investigating the phase structure of ${\eusm T}$ are likely 
to be acquired.

Following such rationale, we propose that vSChSB/confinement 
structure of ${\eusm T}$ may be inferred from low-energy behavior of Dirac 
spectral density $\rho(\lambda)$ in these theories. Gauge invariant 
$\rho(\lambda)$ is thus viewed as a vacuum object already known to encode vSChSB 
via its ``infinitely infrared'' behavior ($\rho(\lambda \!\to\! 0)$ in infinite 
volume) and Banks-Casher equivalence~\cite{Ban80A}. The suggestion is that 
properties of $\rho(\lambda)$ in {\em finite} infrared regime reflect
vSChSB/confinement combination as shown in Fig.\ref{fig:illus1}: case (A) 
corresponds to vSChSB with confinement, (B) to vSChSB without confinement, 
and (C,C') to valence chiral symmetry and no confinement. Note that option (A) 
represents generic behavior observed in low-temperature ``real-world'' QCD 
simulations, and (C,C') are standard possibilities for chirally symmetric 
vacuum.\footnote{Case (C') involving strict ``gap'' in density is very difficult 
to infer from any practical simulation and may or may not exist in ${\eusm T}$. 
However, the distinction between (C) and (C') is immaterial for our purposes.} 
We refer to the atypical case (B) as {\em anomalous}. There are few explanatory 
comments to make.

{\em (i)} Dependencies of Fig.\ref{fig:illus1} refer to limits approached in 
asymptotically large volumes. Non-monotonicities arising e.g. in small volumes 
at fixed topological charge are not invoked. 

{\em (ii)} While Banks-Casher relation is a kinematical constraint, 
the above connection between vacuum properties and $\rho(\lambda)$ is dynamical, 
arising due to the nature of strong~force.

{\em (iii)} Note that confinement without vSChSB is not among the possibilities 
in ${\eusm T}$, aligning with conclusions of Ref.~\cite{Cas79A}.

{\em (iv)} We implicitly assume a fully chiral lattice definition of valence 
quark dynamics via overlap Dirac operator~\cite{Neu98BA}. This makes vSChSB  
and the proposed classification of phases sharply defined at the regularized 
level. Note that the nature of phase identification is such that it is 
immaterial whether bare or renormalized~\cite{Del05A} spectral density is 
used. It is equally immaterial that $\rho(\lambda \!\to\! 0)$ (valence 
chiral condensate) may diverge even at the regularized level in some 
theories~\cite{Kis01A}: this just means that the system is necessarily
in phase (B).

{\em (v)} Bimodality in $\rho(\lambda)$ was first observed on N$_f$=0 
backgrounds~\cite{Edw99A} above deconfinement temperature $T_c$. However, 
its cutoff dependencies have not been systematically studied amidst 
suspicions (see e.g.~\cite{Bui08A}) that it may be a lattice artifact. 
Consequently, substantiating the existence of anomalous phase in ${\eusm T}$ 
is central to our discussion. 

{\em (vi)} Recent observation of anomalous $\rho(\lambda)$ in zero-temperature 
dynamics with many light flavors~\cite{Ale14A,Ale14C} provided an important 
impetus for current generalization: since both thermal and light-quark effects 
induce anomalous phase, it is reasonable to expect its occurrence along generic 
paths in ${\eusm T}$ leading to valence chiral symmetry restoration. 

{\em (vii)} In the current context, the only role of light valence quarks is 
to probe the vacuum: while not influencing it in any way, their infrared 
dynamics is sensitive to its long-range properties. Contrasted with type (A), 
the anomalous density (B) then signifies a different mechanism for 
Goldstone-like correlations, relevant when confinement is absent.

In what follows, we first discuss new results from lattice QCD needed for this 
proposal, i.e. points $(\alpha)$-$(\gamma)$ outlined in the Abstract. We then 
proceed to generalize (Fig.~\ref{fig:Tsets}), taking into account point 
$(\delta)$, and suggest the existence of anomalous phase preceding 
the conformal window. Selected aspects of our results, conclusions and their 
consequences, are elaborated upon in the last section.

\vspace*{0.10in}

\noindent{\bf 2. Reality of the Anomalous Phase.}
Basic requirement for viability of the proposed general scenario is establishing
the existence of deconfined dynamics with vSChSB and anomalous spectral density
anywhere in ${\eusm T}$. Thermal N$_f$=0 theory is ideal for this purpose since 
all three elements involved are well-defined and lattice-accessible. Working with 
Wilson's lattice theory at T/T$_c$=1.12 (fixed by $r_0 T_c$~\cite{Nec03A}), where 
we observed bimodality in spectral density previously~\cite{Ale12D,Ale14A}, 
our aim is to perform crucial stability tests with respect to infrared and 
ultraviolet cutoffs. Our preliminary work in this direction was reported
in Ref.~\cite{Ale14B}. The results below refer to Z$_3$-broken vacuum with 
``real Polyakov line'' \cite{Cha95A}, which gets selected when deforming 
the theory away from N$_f$=0. 

Overlap Dirac operator ($\rho\!=\!26/19$) constructed from its Wilson counterpart 
($r\!=\!1$) is used to define valence quark dynamics throughout this work. Spectral 
density $\rho(\lambda)$ is the right derivative of cumulative function 
$\sigma(\lambda) \!\equiv\! \langle \, \sum_{0<\lambda_i<\lambda} 1 \,\rangle/V$, 
where $V$ is the 4-volume and $\lambda_i$ (real numbers) have magnitudes of Dirac
eigenvalues and signs of their imaginary parts. Indicated exclusion of exact zeromodes 
from counting is harmless since their effect vanishes in the infinite-volume limit. 
Given the finite statistics of any simulation, coarse-graining of $\rho(\lambda)$ 
is unavoidable. In the absence of suspicion for singularity,
we use the symmetric definition away from origin, namely 
\begin{equation}
    \rho(\lambda,\delta) \equiv 
    \frac{\sigma(\lambda+\delta/2)-\sigma(\lambda-\delta/2)}{\delta}
    \qquad\quad \lambda \ge \delta/2 
\end{equation}
When estimating the infinite-volume value of $\rho(\lambda \!\to\! 0)$ from finite 
systems, it is convenient to work with ``right derivative'' form, i.e. 
$\rho(\lambda\!=\!0,\Delta) \equiv \sigma(\Delta)/\Delta$. 
Valence chiral condensate with overlap is identical to
$ \pi \lim_{\lambda\to 0} \lim_{V \to \infty} \rho(\lambda,V)$, 
as is formally in the continuum.

\begin{figure}[t]
\begin{center}
    \centerline{
    \hskip 0.00in
    \includegraphics[width=5.6truecm,angle=0]{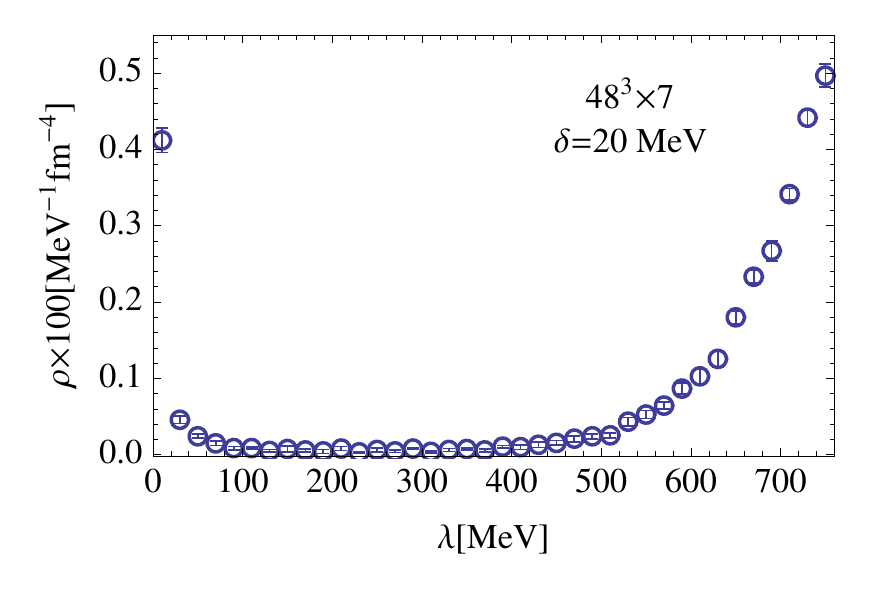}
    \hskip -0.13in
    \includegraphics[width=5.6truecm,angle=0]{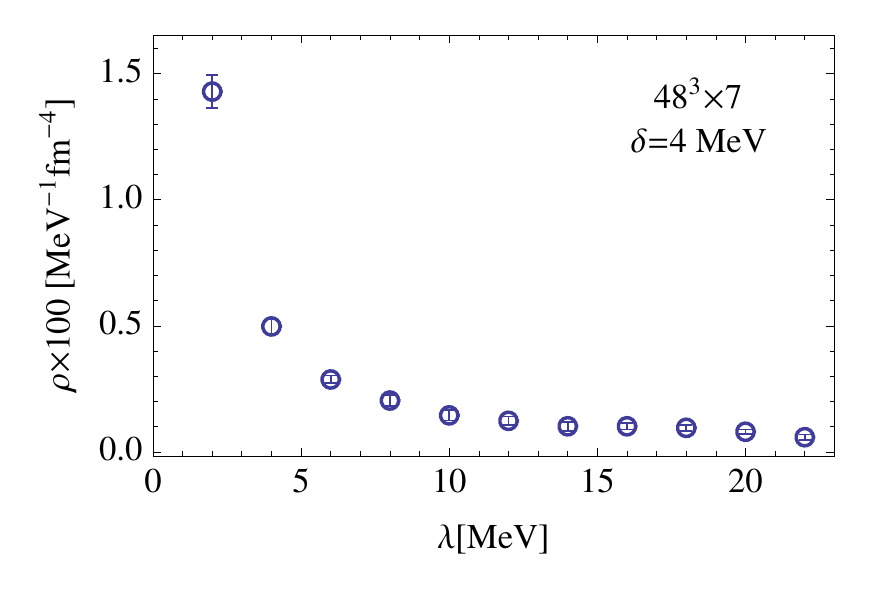}
    \hskip -0.13in
    \includegraphics[width=5.6truecm,angle=0]{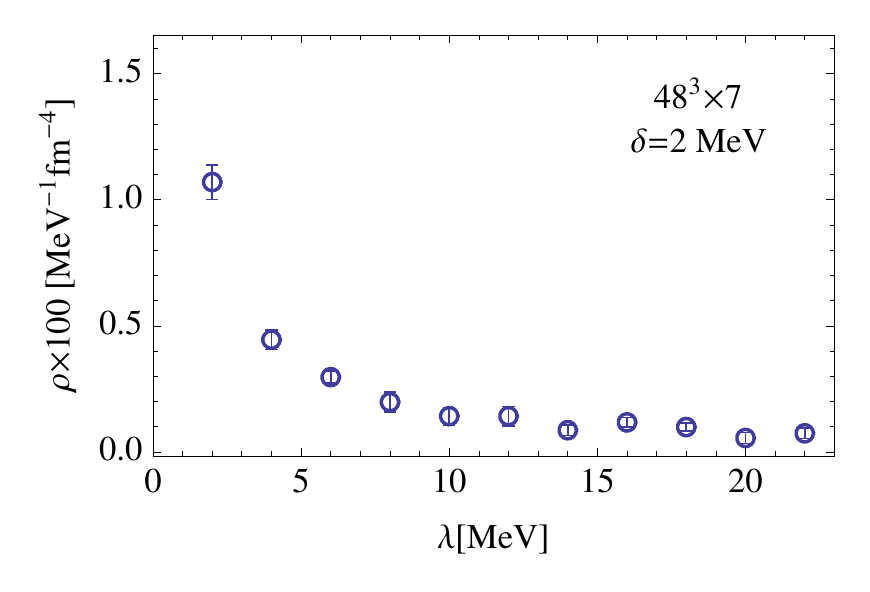}
     }
     \vskip -0.1in
    \centerline{
    \hskip 0.00in
    \includegraphics[width=5.6truecm,angle=0]{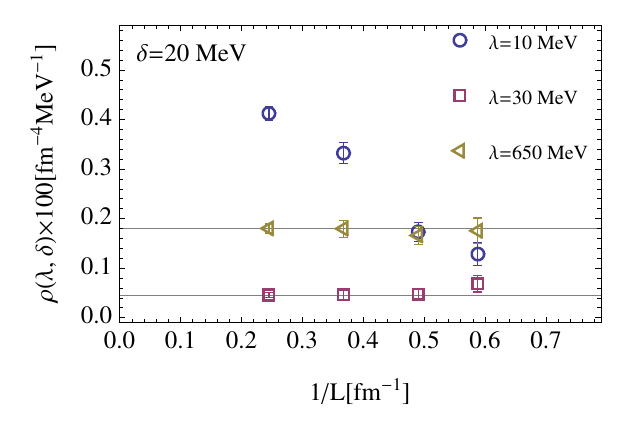}
    \hskip -0.13in
    \includegraphics[width=5.6truecm,angle=0]{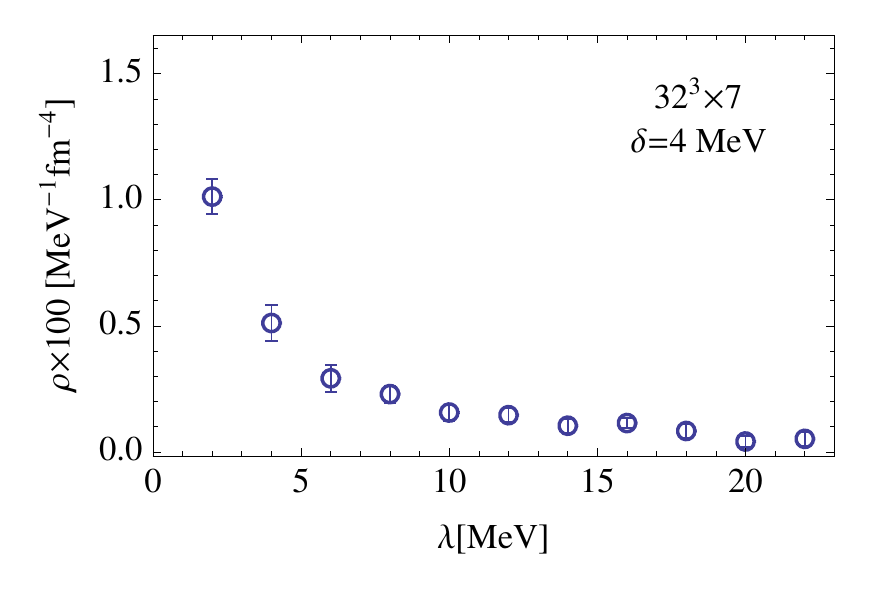}
    \hskip -0.13in
    \includegraphics[width=5.6truecm,angle=0]{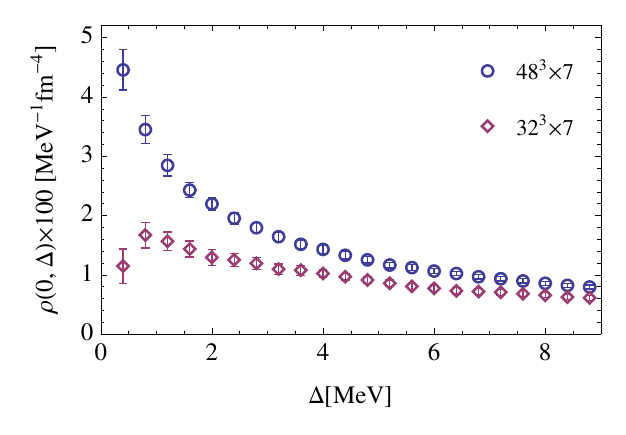}
     }
     \vskip -0.1in
     \caption{Anomalous phase in N$_f$=0 theory at $T\!=\!1.12T_c$ and lattice spacing 
     $a\!=\!0.085\,\mbox{\rm fm}$.}
     \label{fig:infrared}
    \vskip -0.40in
\end{center}
\end{figure} 

Focusing first on the infrared, we simulated $N^3 \!\times\! 7$ systems at 
$N\!=\!20,24,32,48$. To set $T\!=\!1.12 T_c$ requires lattice spacing 
$a\!=\!0.085 \, \mbox{\rm fm}$ 
($\beta\!=\!6.054 \,,\, r_0 \!=\! 0.5 \, \mbox{\rm fm}$), 
giving linear size $L\!=\!4.1 \,\mbox{\rm fm}$ to the largest system. 
Low-lying overlap eigensystems were computed and bimodal $\rho(\lambda)$, such as 
in the top left plot of Fig.~\ref{fig:infrared}, was found at all volumes. 
The spectrum exhibits excellent volume scaling, exemplified in the bottom left 
plot\footnote{The horizontal lines guiding the eye are fits to a constant from 
three largest systems.}, except for the most infrared point, which follows 
a growing trend ensuring the anomalous shape (B) in the infinite-volume limit. 
The coarse-graining $\delta\!=\!20$ MeV is sufficiently fine except for the most 
infrared bin $(0,20) \mbox{\rm MeV}$. Here $\rho(\lambda)$ changes rapidly, 
as the closeup with $\delta\!=\!4$ MeV (top middle) shows. Further lowering of 
$\delta$ (top right) only affects the most infrared point, and we have thus arrived 
at the resolution revealing the shape of the anomalous peak. Importantly, this 
shape is also stable under the change of infrared cutoff: comparing $N\!=\!32$ to
$N\!=\!48$ (lower middle vs top middle), only accumulation in the lowest bin 
changes, growing again. We conclude that, in the infinite volume, $\rho(\lambda)$ 
has a positive (possibly infinite) local maximum at $\lambda\!=\!0$, following 
behavior (B), with anomalous peak of {\em finite width} (few MeV).

\begin{figure}[t]
\begin{center}
    \centerline{
    \hskip 0.00in
    \includegraphics[width=5.6truecm,angle=0]{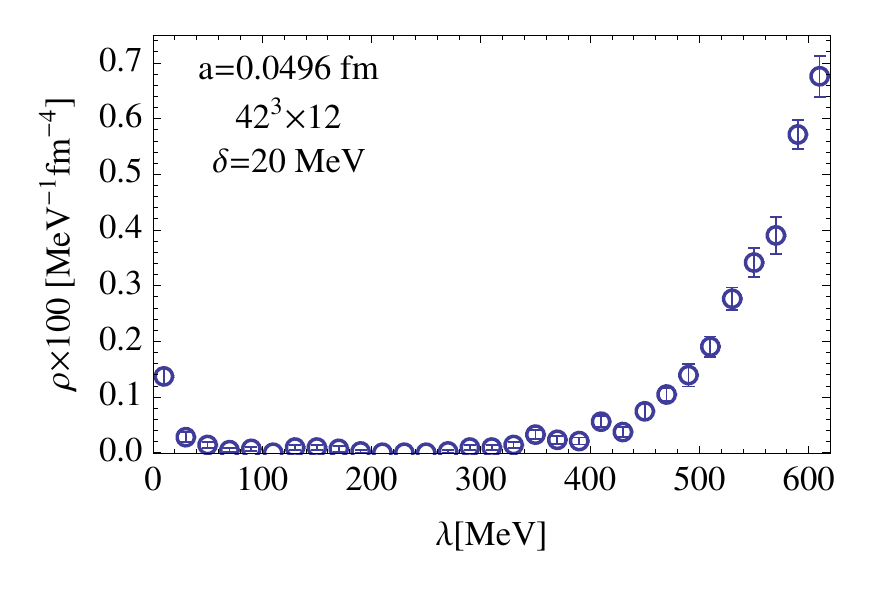}
    \hskip -0.13in
    \includegraphics[width=5.6truecm,angle=0]{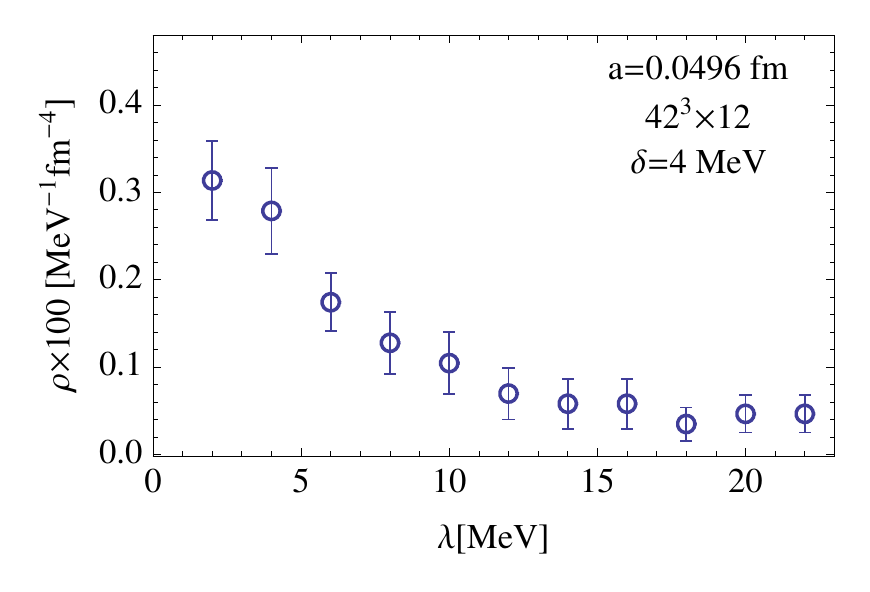}
    \hskip -0.13in
    \raisebox{5pt}{\includegraphics[width=5.7truecm,angle=0]{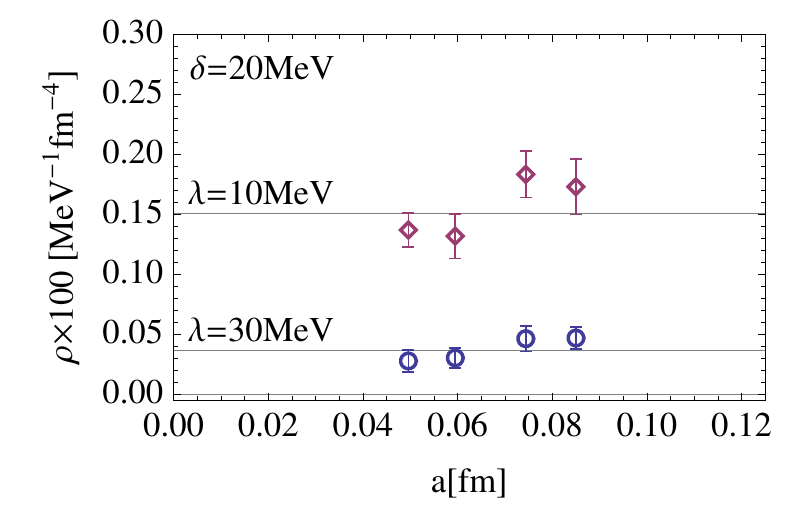}}
     }
     \vskip -0.05in
    \centerline{
    \hskip 0.00in
    \includegraphics[width=5.6truecm,angle=0]{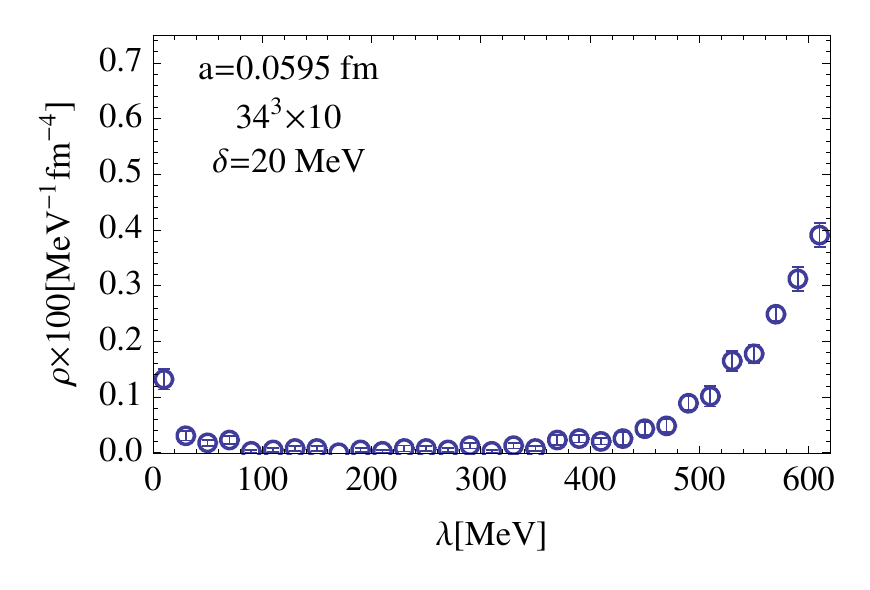}
    \hskip -0.13in
    \includegraphics[width=5.6truecm,angle=0]{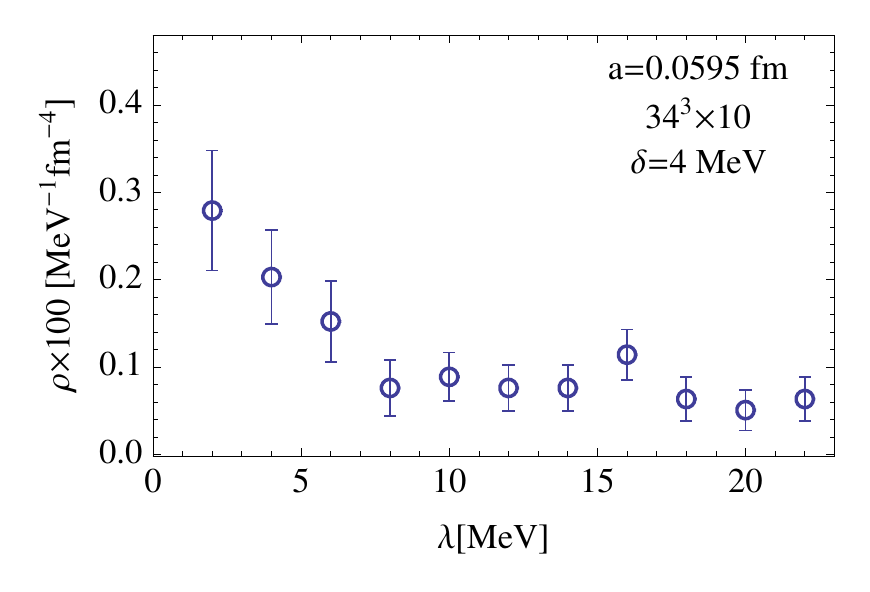}
    \hskip -0.13in
    \raisebox{4pt}{\includegraphics[width=5.7truecm,angle=0]{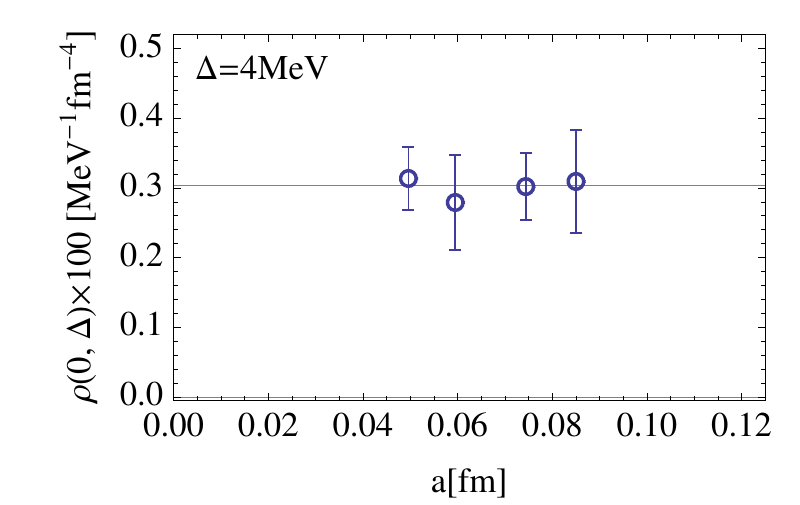}}
     }
     \vskip -0.1in
     \caption{Ultraviolet cutoff and anomalous behavior: N$_f$=0 theory at $T\!=\!1.12\, T_c$, 
     $L\!=\!2.0 \,\mbox{\rm fm}$.}
     \label{fig:conf_ultraviolet}
    \vskip -0.40in
\end{center}
\end{figure} 

While anomalous behavior (B) alone implies it, we address valence chiral symmetry breaking 
explicitly (lower right 
plot of Fig.~\ref{fig:infrared}) via $\Delta$-dependence of $\rho(\lambda\!=\!0,\Delta)$. 
Note that, in any finite volume, $\lim_{\Delta\to 0}\rho(0,\Delta,V)\!=\!0$, 
and the associated downturn, shown for $N\!=\!32$, will occur at sufficiently small $\Delta$. 
However, as expected in broken theory, the break is pushed toward zero with increasing volume. 
Moreover, the estimates at any fixed $\Delta$ are growing, making vSChSB all but certain. 
At $T\!=\!1.12 T_c$ the system is deconfined, and thus all elements of the proposed connection 
are in place for this case.   

Turning to ultraviolet cutoff, we fixed the physical volume of $N\!=\!24$ theory  
($L\!=\!2.04 \,\mbox{\rm fm}$), and drove the system toward continuum limit, while keeping 
$T/T_c\!=\!1.12$. This resulted in sequence 
$N \!\times\! N_t \!=\! 24 \!\times\! 7,\, 28 \!\times\! 8,\, 
34 \!\times\! 10,\, 42 \!\times\! 12$ 
at $a\!=\!0.0850, 0.0744, 0.0595, 0.0496 \,\mbox{\rm fm}$ respectively. The choice of 
moderate physical volume was chiefly motivated by computational considerations. While data
presented in Fig.~\ref{fig:infrared} indicate volume effects in the infrared, the qualitative 
behavior is still very clearly anomalous at this volume, and the setup allows us to explore 
the system rather deep into the continuum limit which is essential.

The bimodal $\rho(\lambda)$ was indeed found at all cutoffs, with global view for the system 
closest to the continuum displayed in Fig.~\ref{fig:conf_ultraviolet} (top left). 
The associated closeup with proper resolution is also shown (top middle). Note that, 
contrary to changing the volume, varying ultraviolet cutoff is not expected to affect very 
infrared scales significantly, unless a qualitative change of dynamics (phase transition) 
occurs, separating lattice and continuum-like behaviors. It is decisively the former 
scenario that is observed, and illustrated by comparisons to the situation at smaller 
cutoff (lower left and middle plots). The scaling of first and second data point at 
$\delta\!=\!20 \, \mbox{\rm MeV}$ (top right) indicates that the bimodal shape of 
$\rho(\lambda)$ will be preserved in the continuum limit at this resolution. Moreover,
the accumulation of very near-zeromodes, i.e. $\rho(0,\Delta)$ with resolution well within 
the natural width of the anomalous peak ($\Delta\!=\!4 \, \mbox{\rm MeV}$), is stable with
cutoff (lower right plot), offering no hint of a qualitative change.

Given the demonstrated dramatic volume effects leading to anomalous phase (B) at physically
relevant cutoff, and the stability of bimodality at fixed volume under ultraviolet cutoff, 
we conclude that {\em continuum} N$_f$=0 theory at $T/T_c\!=\!1.12$ is in the anomalous phase.

\vspace*{0.10in}

\noindent{\bf 3. Confinement and Anomalous Spectral Density.} 
An important aspect of the proposed general scheme is that the transition from confined 
to deconfined theory in ${\eusm T}$ {\em coincides} with the transition from regular 
type (A) of $\rho(\lambda)$ to anomalous type (B). To show this for N$_f$=0 thermal transition, 
we selected a system just below $T_c$. Replacing $N_t\!=\!7$ of our volume study with 
$N_t\!=\!8$ (keeping $\beta\!=\!6.054$, $a\!=\!0.085 \, \mbox{\rm fm}$) corresponds to 
temperature $T\!=\!0.98 T_c$. This nominal assignment uses universal continuum value
$r_0 T_c \!=\! 0.7498(50)$ of Ref.~\cite{Nec03A}, and $T/T_c$ at finite cutoff is expected 
to shift slightly upwards. We thus checked the behavior of Polyakov loop in large volumes 
up to $N\!=\!48$, and confirmed that the system at hand is indeed confined and at the very 
edge of the transition.

\begin{figure}[t]
\begin{center}
    \centerline{
    \hskip 0.00in
    \includegraphics[width=5.8truecm,angle=0]{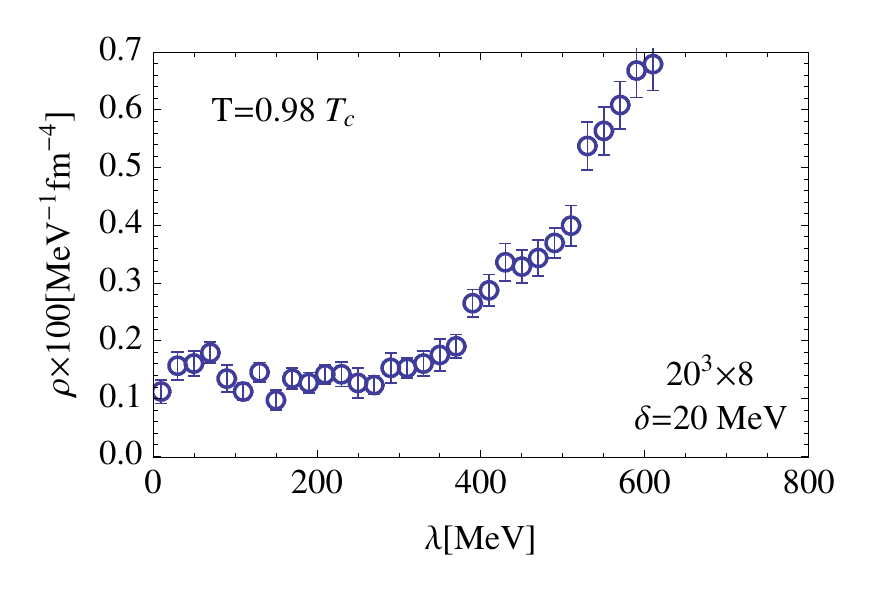}
    \hskip -0.13in
    \includegraphics[width=5.8truecm,angle=0]{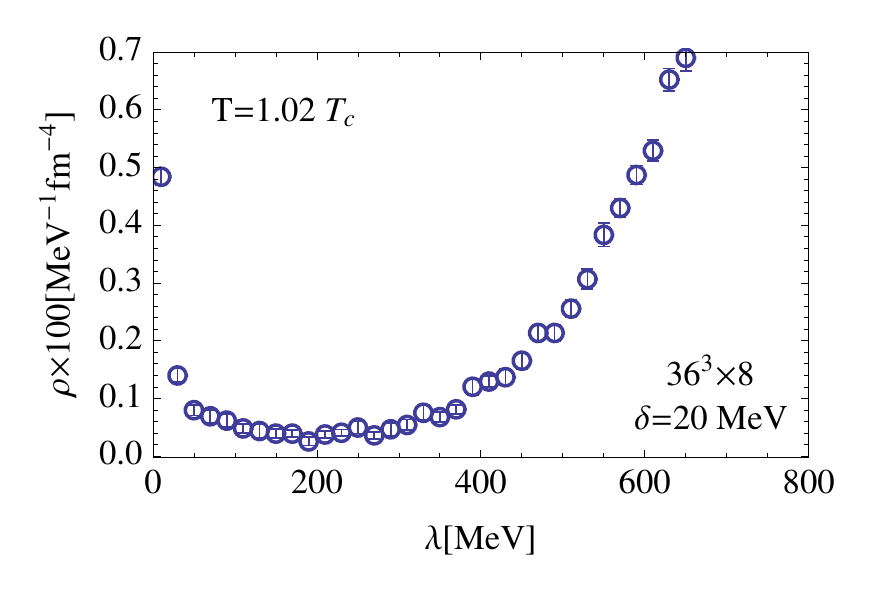}
    \hskip -0.13in
    \includegraphics[width=5.8truecm,angle=0]{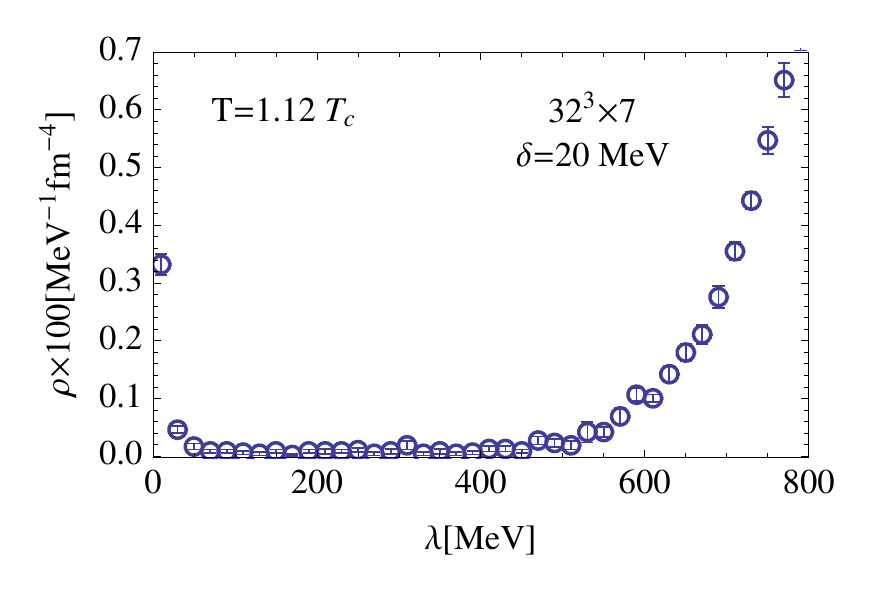}
     }
     \vskip -0.1in
     \caption{Thermal transitions to deconfined and anomalous phase coincide in N$_f$=0 theory.}
     \label{fig:conf_anom}
    \vskip -0.40in
\end{center}
\end{figure} 

Fig.~\ref{fig:conf_anom} (left) shows spectral density for theory so tuned, on 
$20^3\!\times\!8$ lattice. Type (A) behavior is found, as predicted, with flat dependence 
at low energies indicating a system on the verge of the transition. To compare this 
to the situation at temperature just above deconfinement point, we increased the coupling 
slightly ($\beta\!=\!6.0783$, $a\!=\!0.0817 \, \mbox{\rm fm}$) to set $T\!=\!1.02 T_c$. 
Spectral density for such theory on $36^3\times 8$ lattice (middle) shows a dramatic change 
to sharp bimodal behavior: there is little doubt that the Z$_3$ deconfinement transition 
coincides with the transition to anomalous phase. Note that we also show the result for 
system at $T\!=\!1.12T_c$ ($32^3\times 7$, $\beta\!=\!6.054$) in a volume comparable 
to $T\!=\!1.02 T_c$ case, exemplifying the situation well inside the anomalous 
region (right). 

\vspace*{0.10in}

\noindent{\bf 4. Thermal Anomalous Phase of ``Real World'' QCD.} The above evidence 
of anomalous phase in N$_f$=0 QCD, and its association with deconfinement, gives an
essential impetus for its existence in other corners of ${\eusm T}$. It is of particular 
interest to clarify whether anomalous dynamics is part of physical reality for nature's 
quarks and gluons. We address this via N$_f$=2+1 QCD at {\em physical} quark masses, 
i.e. light mass of $(m_u + m_d)/2$ and heavy of $m_s$, since it is well established 
that this theory provides for very precise representation of relevant strongly interacting 
physics. Among other things, calculations in this context led to the conclusion that 
thermal transition of strong interactions is a crossover~\cite{Aok06A}.

To probe the crossover region, we utilize N$_t$=8 lattice ensembles of Wuppertal-Budapest 
group (see~\cite{Bor10A} and references therein), used in precise determination of 
transition temperatures. On a technical side, the simulated theory involves 
Symanzik-improved (tree level) gauge action and stout-improved staggered fermions. 
The physical
line of constant physics was defined via fixing $f_K/m_{\pi}$, $f_K/m_K$ and both scaling  
violations and taste-splitting staggered fermion effects were shown to be small at gauge 
couplings involved. Due to its crossover nature, transition temperature is not a unique 
concept and depends both on the observable used and the defining condition chosen. 
For reference, the temperatures associated with inflection points of (light) scalar
density (``condensate'') and Polyakov line were reported as 
$T_c \approx 155 \, \mbox{\rm MeV}$ and $170 \, \mbox{\rm MeV}$ respectively. Given 
that, we selected ensembles at temperatures $T\!=\!150,\,175$ and 200 MeV, to be close 
to both ends of the crossover, and to examine the possible presence of anomalous phase 
past the transition region.

\begin{figure}[t]
\begin{center}
    \centerline{
    \hskip 0.00in
    \includegraphics[width=5.8truecm,angle=0]{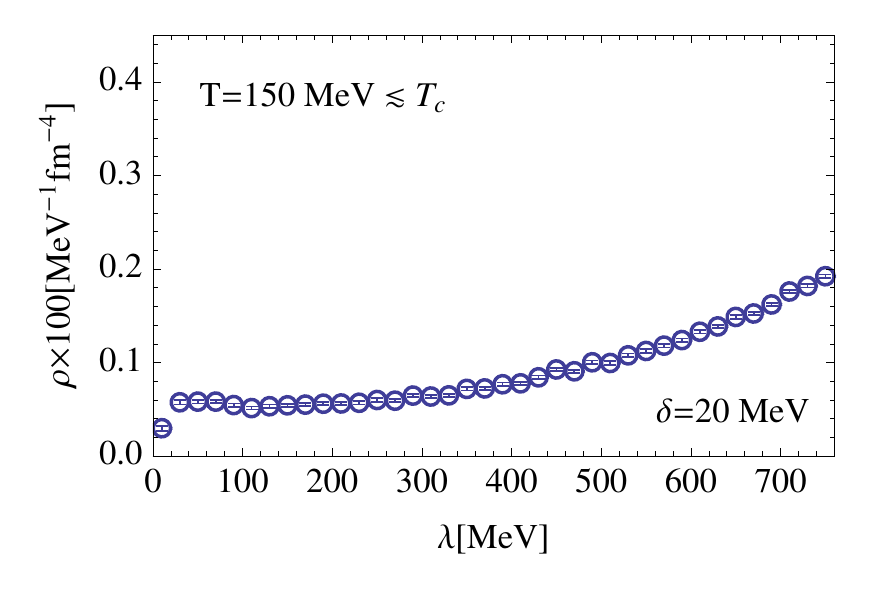}
    \hskip -0.13in
    \includegraphics[width=5.8truecm,angle=0]{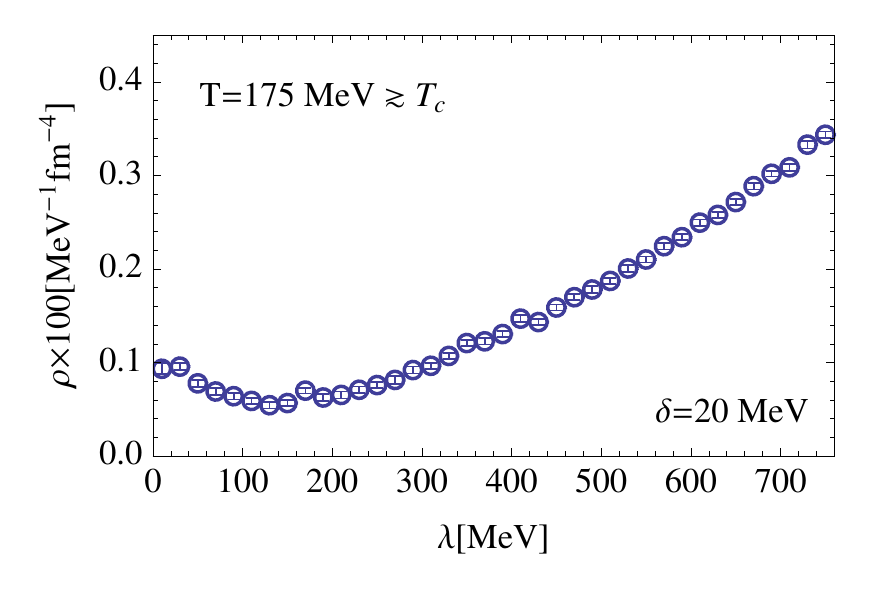}
    \hskip -0.13in
    \includegraphics[width=5.8truecm,angle=0]{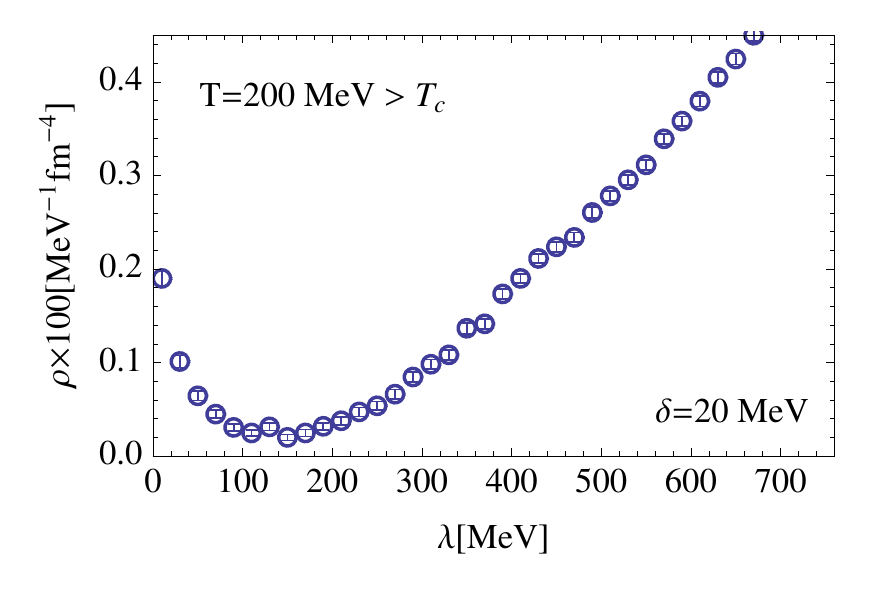}
     }
     \vskip -0.1in
     \caption{Anomalous phase across the thermal crossover in N$_f$=2+1 QCD at physical 
     point.}
     \label{fig:nf2p1}
    \vskip -0.40in
\end{center}
\end{figure}

\begin{figure}[t]
\begin{center}
    \centerline{
    \hskip 0.00in
    \includegraphics[width=6.8truecm,angle=0]{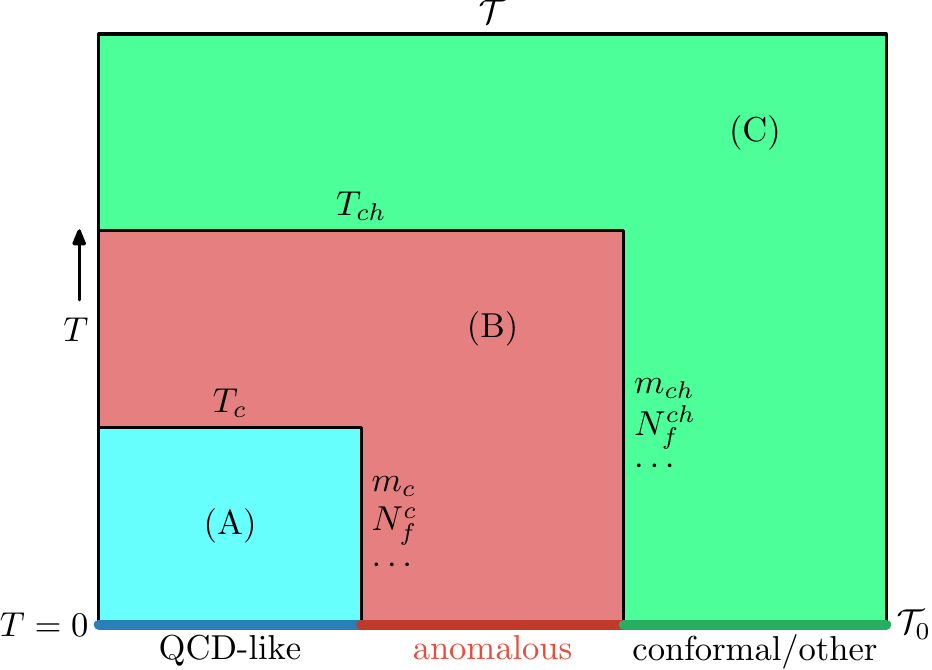}
    \hskip 0.50in
    \includegraphics[width=6.8truecm,angle=0]{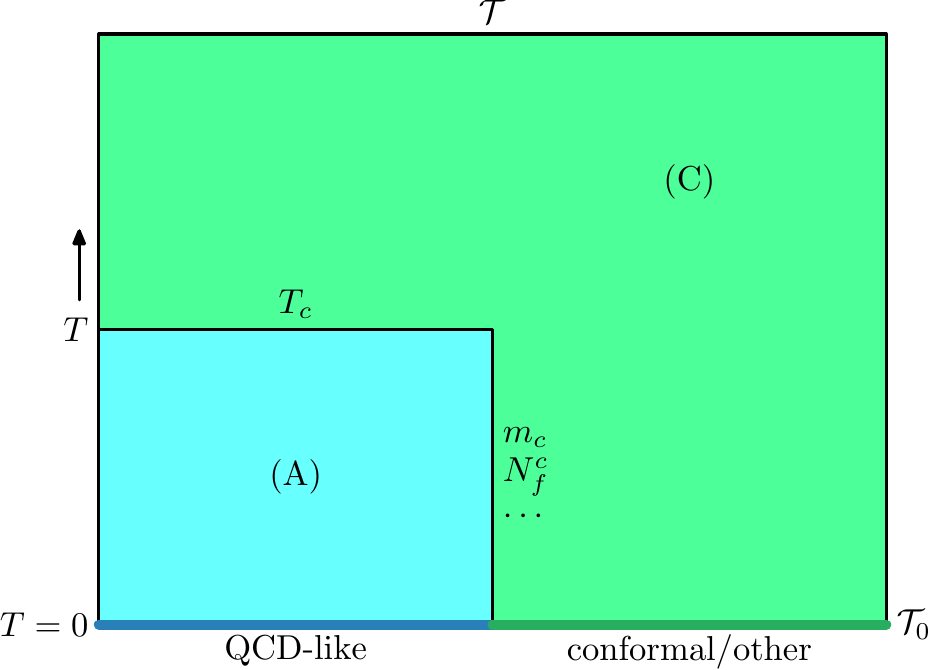}
     }
     \vskip -0.1in
     \caption{The structure of set ${\eusm T}$ proposed here (left) and 
     the conventional one (right).}
     \label{fig:Tsets}
    \vskip -0.35in
\end{center}
\end{figure}

The resulting overlap spectra are shown in Fig.~\ref{fig:nf2p1}. On the lower edge of 
the crossover (left plot), spectral density is extremely flat in the infrared, but 
follows the standard (A) behavior. On the other hand, just past the Polyakov line 
crossover temperature (middle plot), the anomalous behavior clearly sets in. There is 
thus little doubt that the appearance of anomalous phase closely follows the conventional 
measures for expected transition to deconfinement, i.e. conventional ``$T_c$''. 
At $T\!=\!200 \, \mbox{\rm MeV}$, well above the established transition region, 
the anomalous phase fully develops, featuring prominent anomalous peak and the large 
degree of depletion at intermediate scales.

Given the extensive checks performed on the ensembles used, both with respect 
to continuum limit and sufficiency of volumes\footnote{The smallest linear size 
involved here is L=3.9 fm at $T\!=\!200 \, \mbox{\rm MeV}$.} (see~\cite{Bor10A} and 
references therein), we conclude that the anomalous phase $T_c \!<\! T \!<\! T_{ch}$, 
i.e. deconfined phase with broken valence chiral symmetry, exists in high-temperature 
dynamics of strong interactions, and follows the general scheme proposed here.

\vspace*{0.10in}
\noindent{\bf 5. Generic Anomalous Phase, N$_{\mathbf f}^{\mathbf c}$ and 
N$_{\mathbf f}^{\mathbf ch}$.} We have presented evidence that anomalous phase with 
claimed properties {\em exists} in ${\eusm T}$, as well as evidence that thermal 
effects lead to bimodal $\rho(\lambda)$ rather generically. Regarding the latter, 
apart from results presented here, observation of bimodal behavior in SU(N) theories 
was reported under varied but always thermal circumstances, e.g. 
in Refs.~\cite{Edw99A,Kis01A,Bui08A,Ale12D,Cos13A,Buc13A,Sha13A,Ale14A}.  
Important additional ingredient is provided by the recent finding that presence of 
light dynamical quarks, without any thermal agitation, also leads to anomalous phases 
at sufficiently large number of flavors~\cite{Ale14A,Ale14C}. Thermal and light-quark 
effects thus appear to be analogous in this regard. The relevance of this stems from 
the fact that thermal effects (increasing $T$), light-quark effects (decreasing masses, 
increasing $N_f$), and their combinations, are the only available freedoms 
in ${\eusm T}$ to facilitate the transition from broken valence chiral symmetry to 
its restoration. This leads us to propose that anomalous phases commonly occur on 
paths to chirally symmetric vacua. Our evidence on simultaneity of transitions
to deconfinement and anomalous $\rho(\lambda)$ (in known cases) completes the rational 
basis for the proposed picture.

Note that for N$_f$=12 theories, the above scenario was seen~\cite{Ale14A,Ale14C} 
to imply the existence of a mass $m_c$ below which the anomalous phase appears. 
The phase may either extend down to $m\!=\!0$ or to non-zero $m_{ch}\!<\!m_c$, 
namely the possible point of valence chiral restoration. 
Here we wish to point out another special case of the above general argument, 
concerning the ``path'' in ${\eusm T}$ parametrized by the number of massless quark 
species N$_f$ at $T\!=\!0$. Indeed, it is widely believed that the N$_f$=2 case 
represents a confining, chirally broken theory with type (A) spectral density. 
At the same time, since the work of Banks and Zaks~\cite{Ban82A}, it is expected that 
there is a critical number of flavors N$_f^{cr} \!<\! 16.5$, such 
that theories in the window N$_f^{cr} <$ N$_f < 16.5$ are both asymptotically free, and 
controlled at low energy by a conformal infrared fixed point.\footnote{Note that we 
consider N$_f$ to be an integer, while boundary values such as N$_f^{cr}$, N$_f^c$ and 
N$_f^{ch}$ half-integers.} Theories in the conformal window cannot generate low-energy 
scales (at least not below the scale of conformality), and the standard scenario is 
that N$_f^{cr}$ marks the common transition to deconfined, chirally symmetric phase 
with $\rho(\lambda)$ of type~(C).

However, based on the proposed picture, we predict that when increasing the number
of flavors from N$_f$=2, quark-gluon dynamics will first lose confinement 
(at N$_f^c$), and only then chiral symmetry breaking (at N$_f^{ch}$), generating 
the anomalous phase for
\begin{equation}
  2 < \mbox{\rm N}_f^c < \mbox{\rm N}_f < \mbox{\rm N}_f^{ch} = 
  \mbox{\rm N}_f^{cr}
\end{equation}
The last equality holds if the current view associating the onset of conformal 
window with chiral symmetry restoration is valid. It is important in this regard that 
N$_f$=12 may be in the anomalous phase~\cite{Ale14A,Ale14C}. Indeed, the setup with yet 
fewer flavors is even more conducive to the possibility of anomalous vSChSB surviving 
masslessness (SChSB), since less of a condensate-destructive light-quark effect
is generated. N$_f^c$ could thus be quite low.
 
Schematic structure of the set ${\eusm T}$, we are proposing, is shown in 
Fig.~\ref{fig:Tsets} (left). At $T\!=\!0$ (set ${\eusm T}_0$) there are theories with all 
three types of $\rho(\lambda)$ and associated vacuum properties. We refer to type (A) as 
QCD-like for this purpose. Type (C) is expected to contain theories from conformal
window and e.g. theories without asymptotic freedom. Heating up a QCD-like system generically 
involves both deconfinement ($T_c$) and separately valence chiral restoration ($T_{ch}$), 
while anomalous type (B) theory only undergoes the latter, and type (C) neither. Moving 
within ${\eusm T}$ at fixed temperature involves analogous transitions. 
Note that other kinds of changes, unrelated 
to vSChSB and confinement, may occur within the three regions. Also, we do not imply that 
phases (A) and (C) are strictly disconnected, with anomalous phase occurring along every 
path between them: the intention is to convey a typical situation.

\vspace*{0.10in}
\noindent{\bf 6. Important Points.} In this work, we have proposed a specific association 
between infrared behavior of Dirac spectral density, and confinement/valence chiral 
symmetry breaking structure of SU(3) gauge theories with fundamental quarks 
(Fig.~\ref{fig:illus1}). There are few points we wish to discuss and highlight
in this regard.
   
{\bf I.} A significant conceptual step implied by our analysis and represented by 
Fig.~\ref{fig:Tsets}, is that valence chiral symmetry breaking without confinement is 
a commonplace for SU(3) gauge theories with fundamental quarks: the schematic view on 
the left, with many new transitions, is proposed to be a good representation of 
the situation, as opposed to the standard view on the right. The {\em existence} of 
anomalous phase in ${\eusm T}$, shown here, is pivotal for this transformation.

{\bf II.} The proposed association of Dirac spectral density with vacuum properties turns
it into a valuable tool for investigating the phase structure of ${\eusm T}$. Indeed, 
recall that standard methods, such as those used in thermal lattice QCD, are ``transitional'' 
in nature, requiring the study of {\em changes} in the theory space. On the other hand, 
utilizing $\rho(\lambda)$ is as convenient as order parameter at hand: it provides 
a definite answer for any standalone theory. A single-number indicator of anomalous 
phase (``order parameter'') 
can be defined as follows. Let $\lambda_{an}$ be the largest scale such that 
$\rho(\lambda,V \!\to\! \infty)$ strictly decreases on $[0,\lambda_{an})$. Then 
\begin{equation}
    \Omega_{an} \equiv \sigma(\lambda_{an}) > 0 
\end{equation}
i.e. a non-zero volume density of anomalous modes, identifies the theory in anomalous 
phase.\footnote{Note that $\lambda_{an}$ itself is such an indicator.} This definition 
makes $\Omega_{an}$ analogous in nature to $\Omega$, the volume density of chirally 
polarized modes, which was proposed to be a valid order parameter of 
vSChSB~\cite{Ale12D,Ale14A}.

{\bf III.} The importance of concluded anomalous (deconfined with broken valence 
chiral symmetry) phase in physics of quarks and gluons is hard to overstate. Indeed, 
the temperature regime involved is relevant for strongly interacting dynamics associated 
with the creation of plasma-like state currently studied at RHIC and LHC. While 
the bimodal aspect of $\rho(\lambda)$ didn't remain completely unnoticed with regard 
to quark-gluon plasma 
phenomenology~\cite{Kal12A}, the reality of anomalous phase, and the associated details, 
provide strong rationale for detailed exploration of compatible models. We will focus 
on such issues in a separate study.

The emerging evidence for strong interactions generating two ``high-temperature'' 
phases, rather than one, complicates the associated physics in an intriguing way. 
This complexity can be seen in chiral polarization properties of Dirac modes. 
Indeed, our results imply that, past the thermal crossover, quark dynamics generates
rich multilayered chiral structure in Dirac spectrum, shown in  Fig.~17 of 
Ref.~\cite{Ale14A} (left plot).

{\bf IV.} If the proposed anomalous phase separating QCD-like behavior and infrared 
conformality materializes, it is expected to be instrumental for understanding of 
massless quark dynamics, and of transition to conformality in particular. It may also 
complicate some options being considered for realization of the walking technicolor 
scenario~\cite{Yam86A,App86A}. While it is exactly the near-conformal theories that 
are of interest for that purpose, one should emphasize that dynamics in anomalous 
phase is still strongly coupled, and that deconfinement doesn't imply a complete 
absence of hadronic bound states. The theories from ``anomalous window'' could thus 
still be relevant in the context of walking technicolor.

\bigskip

\noindent {\bf Note Added:} Few days before the first version of this manuscript  
was released, the work relevant for point $(\gamma)$ of the Abstract also appeared 
\cite{Dic15A}. While the discussion of \cite{Dic15A} is carried out in the context 
of $U_A(1)$ restoration problem, their results in N$_f$=2+1 theories with light mass 
not far from physical, are consistent with and corroborate our conclusion that thermal 
effects generically lead to the anomalous phase (as defined here).

\vspace*{0.10in}

\noindent{\bf Acknowledgments:} 
We are indebted to Wuppertal-Budapest collaboration, particularly to Zolt\'an Fodor
and Szabolcs Bors\'anyi, for sharing their lattice ensembles. The help of Mingyang 
Sun is gratefully acknowledged. A.A. thanks for the hospitality of the Physics Department 
at the University of Maryland where part of this work was carried out. A.A. is supported 
by U.S. National Science Foundation under CAREER grant PHY-1151648. I.H. acknowledges 
the support by Department of Anesthesiology at the University of Kentucky.

\end{document}
\bye